\documentclass[11pt]{article}
\usepackage{moriond,epsfig,subfigure}

\def\be{\begin{equation}}
\def\ee{\end{equation}}
\def\bea{\begin{eqnarray}}
\def\eea{\end{eqnarray}}

\newcommand{\pbinv}{{\rm pb}^{-1}}
\def\Gcs{\mbox{GeV/$c^2$}}
\def\Gc{\mbox{GeV/$c$}}
\begin{document}
\vspace*{4cm}
\title{NON SUSY SEARCHES AT THE TEVATRON}

\author{ LAURENT DUFLOT \\ for the CDF and D\O\ collaborations }

\address{Laboratoire de l'Acc\'el\'erateur Lin\'eaire,
 \\ Universit\'e Paris-Sud 
 B\^at~200 \\ 91898 Orsay cedex FRANCE}

\maketitle\abstracts{The CDF and D\O\ experiments have collected and analyzed 
 about $300\,\pbinv$ of data during the Run\,II of the Tevatron. Results of 
searches for new non supersymmetric particles based on these datasets will be 
presented.}

\section{Introduction}
Although supersymmetry (SUSY) is one of the most popular models of physics 
beyond the Standard Model, alternative models have been proposed. Models
of additional spatial dimensions are actively developed. Other models 
include technicolor, compositeness and models with additional Z and/or W like 
bosons.

The Tevatron is currently at the high energy frontier with 1.96\,TeV 
center-of-mass energy and thus is one of the major facilities for the 
discovery of
new particles. Searches for new particles at the Tevatron use one or several 
tool(s) to control 
the huge QCD background : presence of electrons and/or muons, missing 
transverse energy (MET), identified heavy flavor jet, identified tau decays, 
etc\ldots  

In the following, the results of searches for new non SUSY particles will be 
briefly described, more details can be found elsewhere~\cite{webcdf,webdzero}.

\section{Leptonic final states}
 Unless explicitly mentioned, the leptonic final states only cover the final
states with electrons and/or muons.

\subsection{Dilepton final states}
The di-electron and di-muon samples have low background contamination
and a well understood Drell-Yan contribution. They form a good sample for
searching for new particles in direct search (particles decaying to di-leptons 
like extra gauge bosons, technicolor particles, Kalusa-Klein excitations in some 
extra dimensions models, \ldots) or in indirect effects in di-lepton production
(compositeness, extra dimension models, \ldots).

 New physics would often be visible only in the high end of the invariant mass 
spectrum. Since electrons are measured in calorimeters, the di-electron sample
benefit from a good mass and transverse momentum (pT) resolution, and easy 
triggering conditions. On the other hand, one has to control and understand the
contamination from QCD events where jets are mis-identified as electrons. This
has to be measured from real data as the simulation is not reliable enough at 
that level of precision. CDF and D\O\ have selected samples based on 
approximately $200\,\pbinv$, selecting events with two electrons with pT in 
excess of $25\,\Gc$, one of which being required to be in the central part of
the calorimeter. Similarly, di-photon samples are selected. The QCD background 
is less important in the isolated di-muon sample, but special care has to be 
taken to understand and reduce the background from cosmic rays. The CDF sample
is based on $\sim\,200\,\pbinv$ and requires two muons with pT of excess of 
$20\,\Gc$, one with $|\eta|<1.0$ and one with $|\eta|<1.5$. The D\O\ sample is 
based on $\sim\,250\,\pbinv$ and requires two muons with pT of excess of 
$15\,\Gc$. A good agreement is found between the data and the prediction from
Standard Model MC complemented by background estimates from real data
(QCD and cosmics).

\begin{figure}
\begin{center}
\begin{tabular}{cc}

\subfigure[Excluded domain in the coupling versus mass of the first excitation
of the graviton in the Randall-Sundrum model of extra dimensions based on a 
di-electron and di-photon sample of 
$\sim\,200\,\pbinv$ of D\O\ data. \label{fig:RSdezro}]{\epsfig{figure=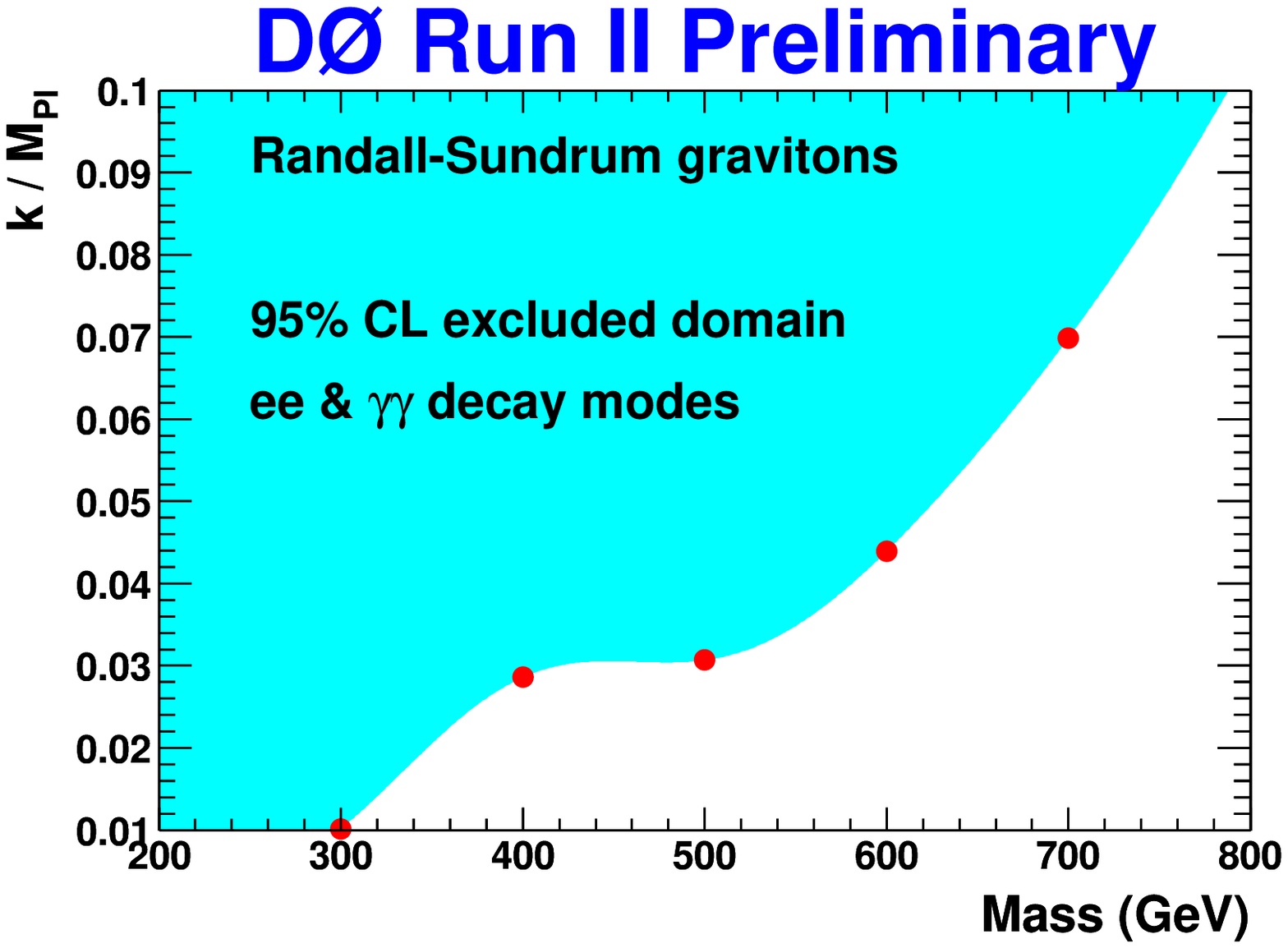,height=4cm}}
&
\subfigure[Cross-section times branching ratio for a sequential Z' boson 
decaying to taus with the upper limit derived from $195\,\pbinv$ of CDF data.
\label{fig:ZpCDF}]{\psfig{figure=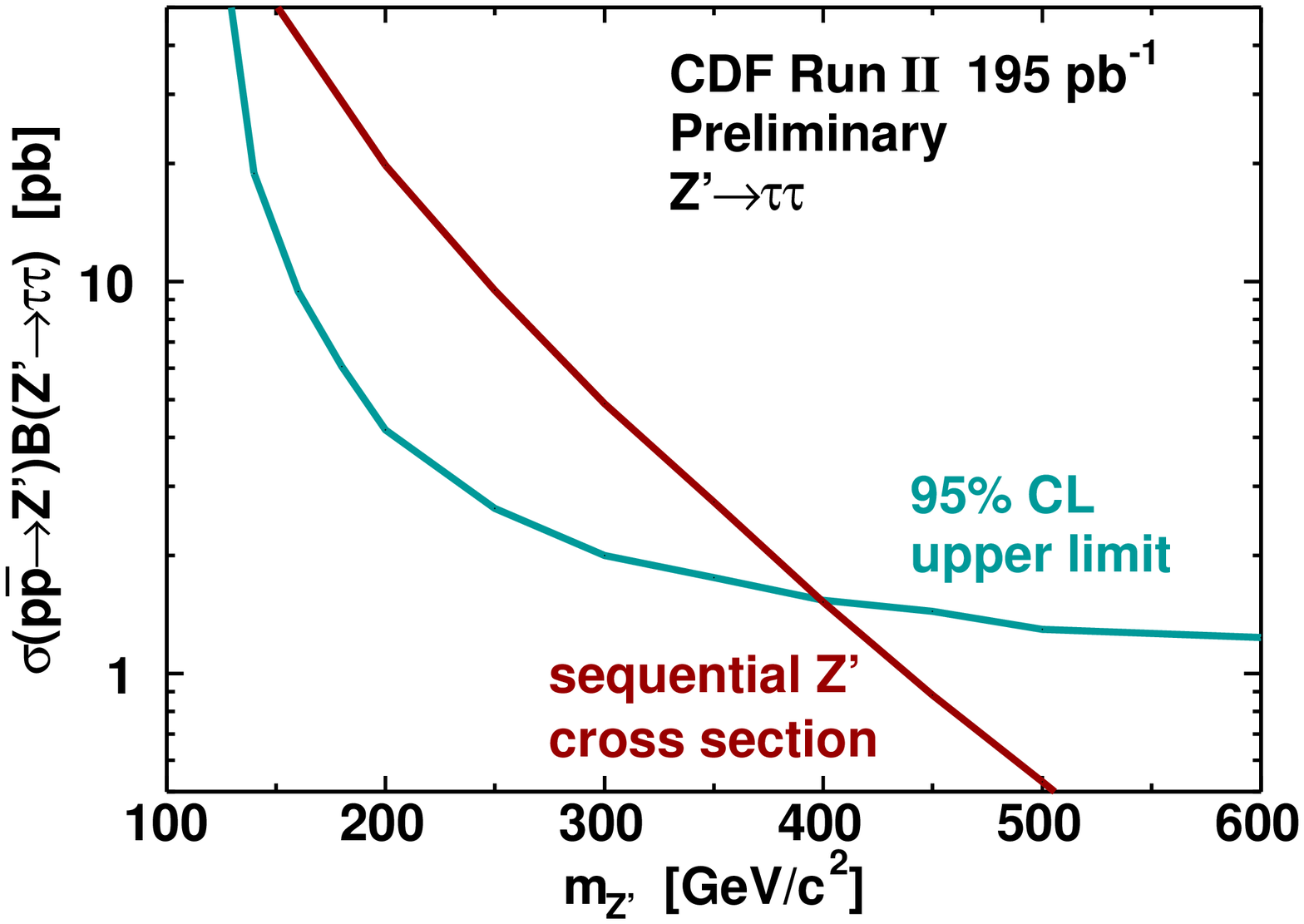,height=4cm}}

\end{tabular}
\end{center}
\caption{Di-lepton results}
\end{figure}

Since no significant deviation from the Standard Model has been observed, the
results are interpreted as limits within various models of new physics. A few
selected results are shown here, additional results can be found 
elsewhere~\cite{webcdf,webdzero}.  Figure~\ref{fig:RSdezro} show the excluded 
domain in the coupling versus mass of the first excitation of the graviton in 
the Randall-Sundrum model of extra dimensions from D\O. An interpretation of 
CDF
results combining the electron and muons channels allow to set a limit on the 
mass of an additional sequential Z boson at $815\,\Gcs$. This search has been
extended to the tau channel as can be seen in figure~\ref{fig:ZpCDF}.


\subsubsection{Search for a Z from long lived parents}
\begin{figure}[h]
\begin{center}
\psfig{figure=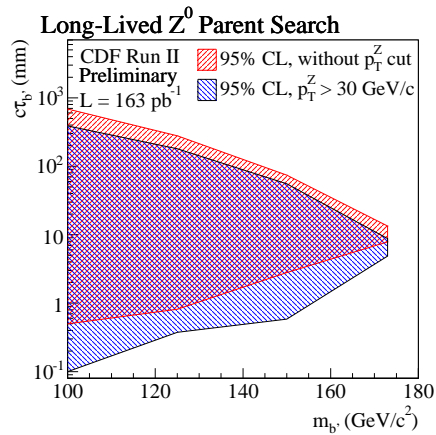,height=6cm}
\end{center}
\caption{ Limits in the decay length versus b' mass from CDF.
\label{fig:ZlonglivedCDF}}
\end{figure}

In some models, Z bosons can be produced in the decays of long lived particles.
CDF looks for such a striking signature in a di-muon sample from $163\,\pbinv$ 
of data. In addition to
selecting the Z mass region, additional criteria are used to ensure a good
reconstruction of the Z vertex, namely tight track quality criteria and the 
requirement of a minimum acoplanarity between the two muons. Two independent 
analyses are then applied : in the first, a tight cut of 0.1\,cm is applied on 
the transverse decay length $L_{xy}$ of the di-muon system (with 2 events 
observed for $0.72\pm0.27$ expected) and, in the second,
a loose
$L_{xy}>0.03\,$cm cut is applied while the di-muon system is required to have
a pT$>30\,\Gc$ (with 3 events observed and $1.1\pm0.8$ expected). The results 
are interpreted in the model of a sequential b' quark lighter than the top 
quark in figure~\ref{fig:ZlonglivedCDF}.

\subsection{Search for a W'}

The di-lepton samples allowed to search for a new Z-like gauge boson. CDF has 
extended that search to that of a W-like boson (W') in the electron channel. 
Events with one isolated electron with pT$>25\,\Gc$,  MET$>25\,$GeV and 
$0.4 < \mathrm{pT}/\mathrm{MET} < 2.5$ (to reject QCD background) are selected.
Assuming Standard Model couplings strength, the analysis result in a limit
on the W' mass of $842\,\Gcs$ using $205\,\pbinv$ of data, 
with systematics dominated by the PDF 
uncertainties and the understanding of the electron energy scale.

\section{Lepton plus jets final states : leptoquarks}
Leptoquarks arise in various models beyond the Standard Model. They decay to 
a lepton and a quark or a neutrino and a quark. First generation leptoquarks are 
supposed to decay only to electron and/or electron neutrino while second 
generation leptoquark would decay only to muons and/or muon neutrinos. The 
branching fraction to the charged lepton is model dependent, so the searches 
are designed to cover all three final states : di-lepton and jets; lepton, MET 
and jets and jets plus MET.

\begin{figure}
\begin{center}
\begin{tabular}{cc}

\subfigure[First generation]{\psfig{figure=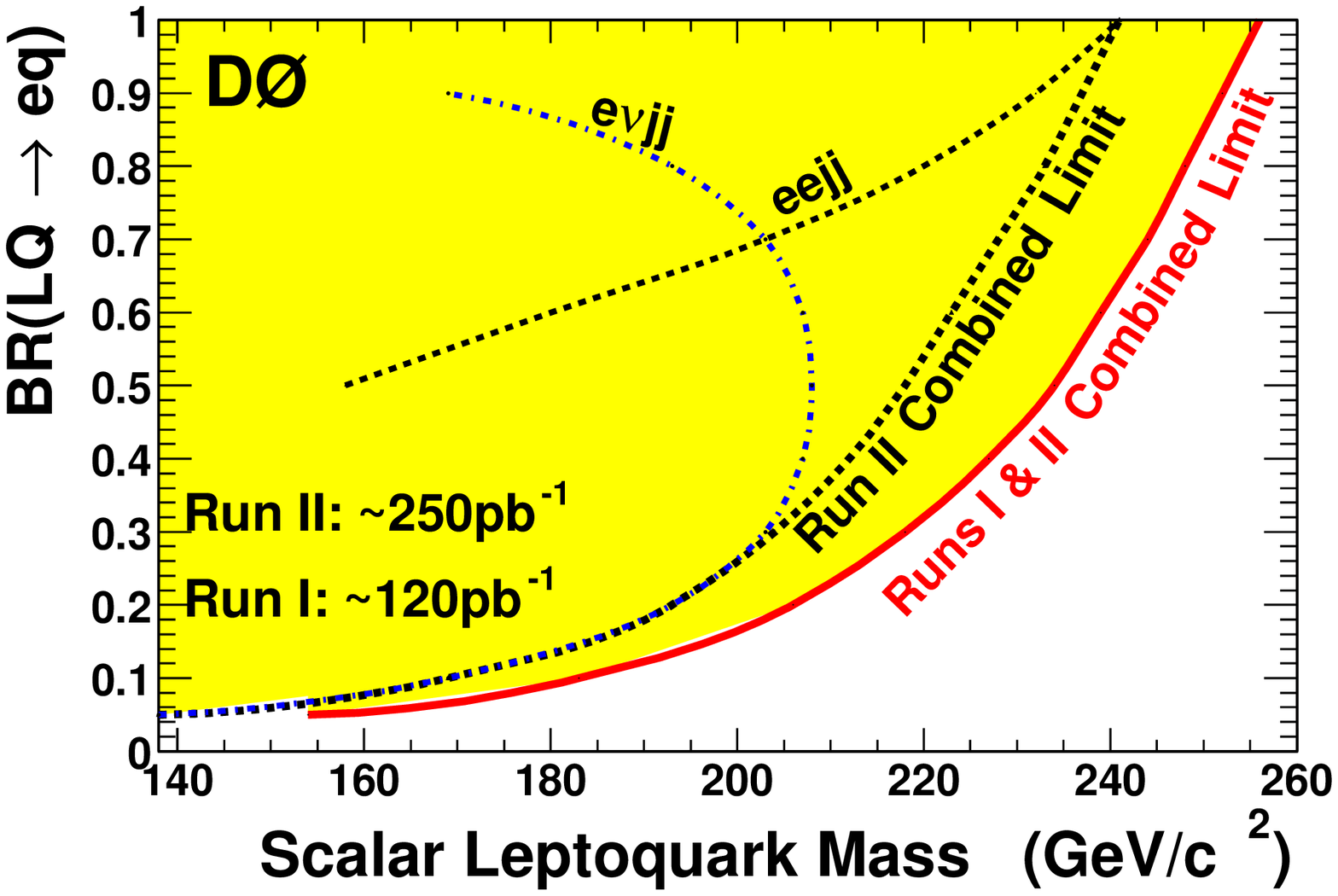,height=5cm}}
&
\subfigure[Second generation]{\psfig{figure=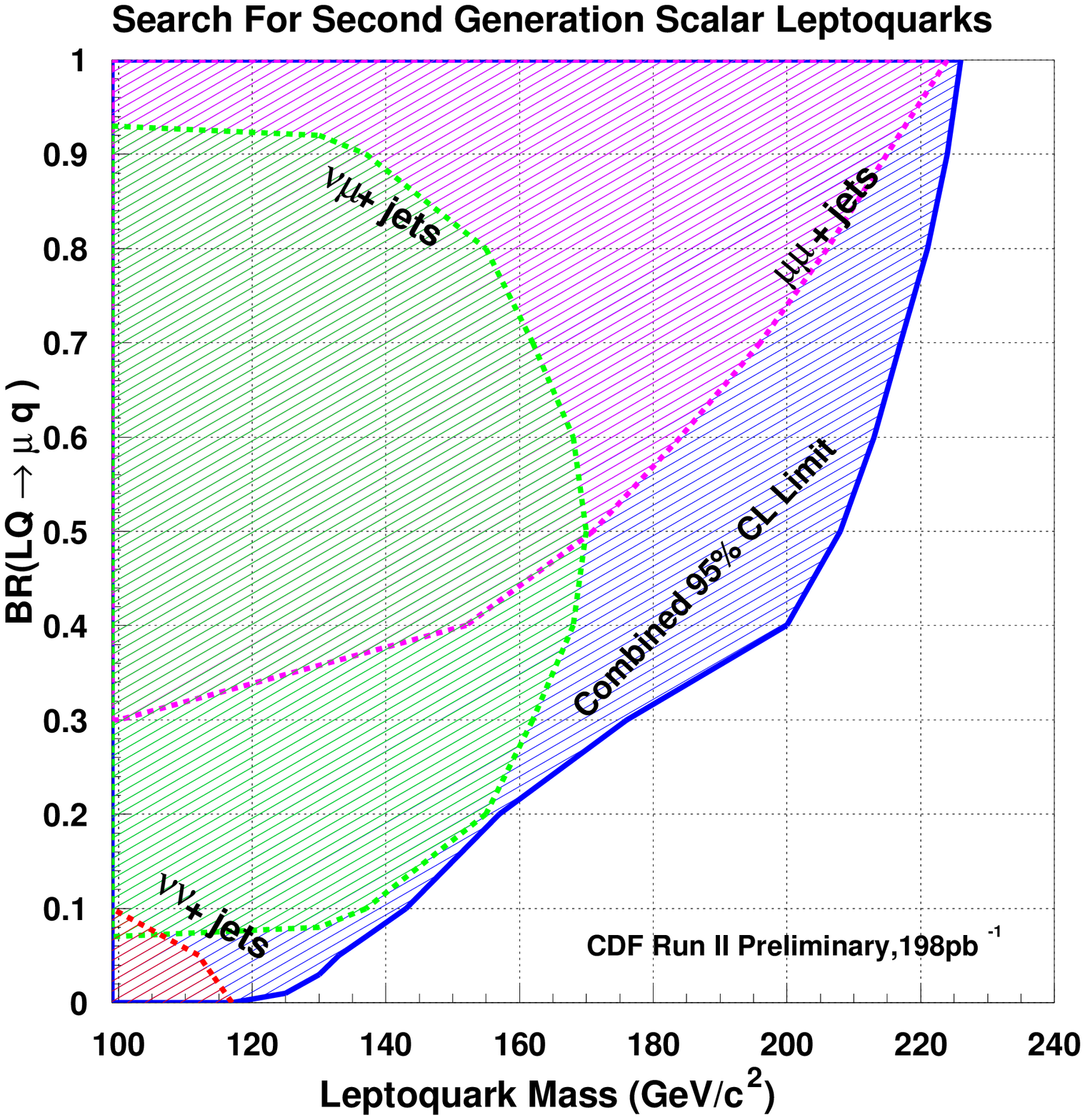,height=5cm}} 

\end{tabular}
\end{center}
\caption{ Limit on the scalar leptoquark mass as a function
of the branching ratio to lepton+quark.
\label{fig:LQ}}
\end{figure}

Limits on first and second generation scalar leptoquarks from D\O\ and CDF are 
shown in figure~\ref{fig:LQ}.

\section{Anomalous W plus heavy flavor production}

\begin{figure}
\begin{center}
\begin{tabular}{ccc}

\subfigure[Lepton tag]{\psfig{figure=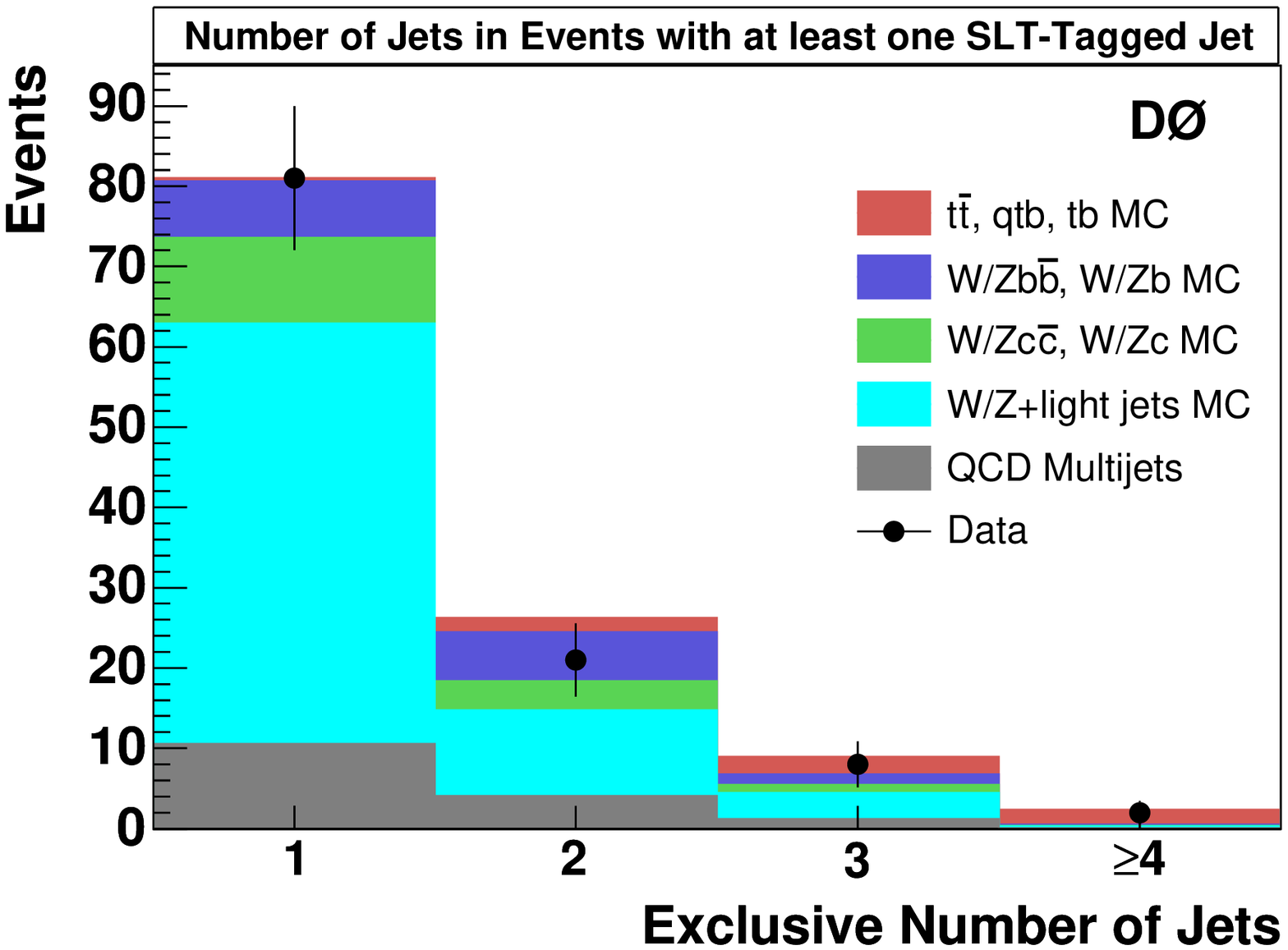,height=3.5cm}}
&
\subfigure[Secondary vertex tag]{\psfig{figure=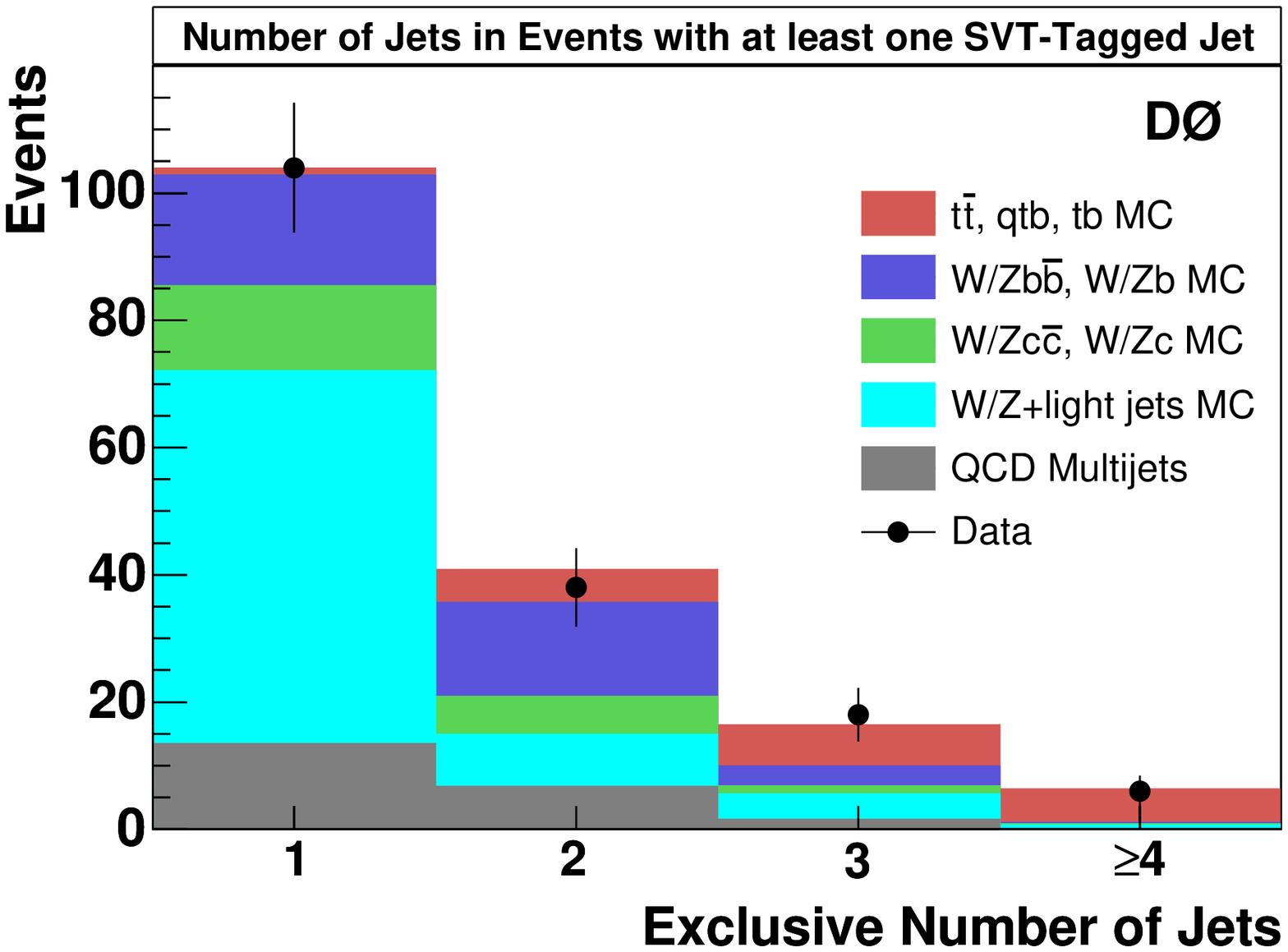,height=3.5cm}}
&
\subfigure[Double tag]{\psfig{figure=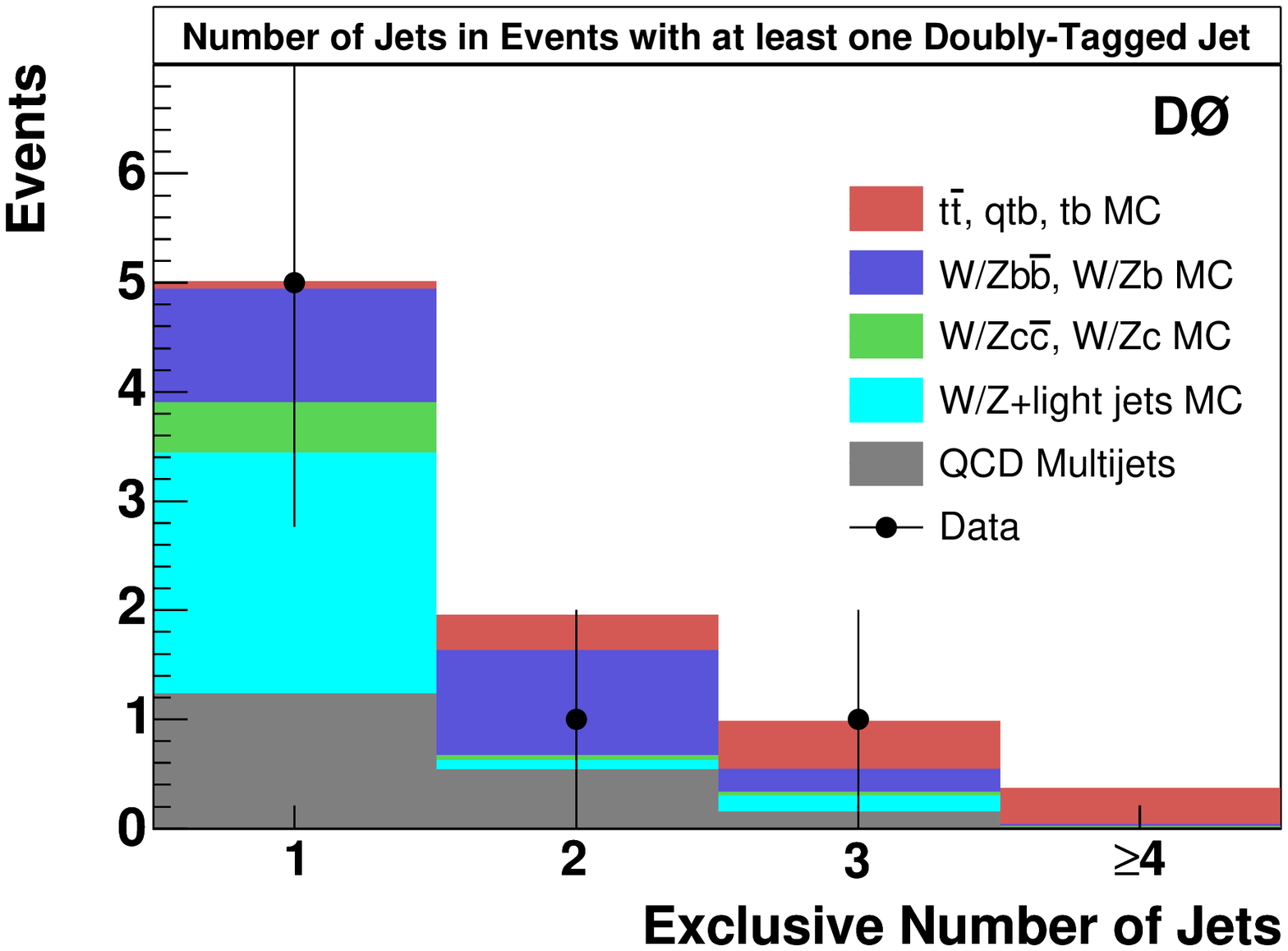,height=3.5cm}}

\end{tabular}
\end{center}
\caption{ Number of jets tagged by one or both of the b-tagging algorithms.
\label{fig:Wbdzero}}
\end{figure}

Studying the associated production of a W with heavy flavor jet(s) is important
as it constitutes a background to e.g. top and Higgs selections. Conversely,
any deviation from Standard Model expectations would be a hint of new physics.
D\O\ has looked for anomalous production of heavy flavor jet in association
with a W in the electron and muons channels using $\sim\,150\,\pbinv$ of data.
The events are required to have one electron (resp. one muon) with 
pT$>20\,\Gc$ and 
$|\eta|<1.1$ (resp. $|\eta|<1.6$), with MET$>20\,GeV$ and not aligned to
the lepton. The transverse mass of the lepton and MET is required to be between
$40\,\Gcs$ and $120\,\Gcs$. Jets are potentially identified as b-jets using
two independent algorithms, one based on the presence of a muon in or nearby 
the jet and the other on the reconstruction of secondary vertices. As can be 
seen in figure~\ref{fig:Wbdzero}, a good agreement is found on the number of 
tagged jets between data and expectation for MC events. As no excess is seen, 
limits are set on the production cross-section of a $Wb\bar b$ type signal
at 26.3\,pb and on a top-like signal at 14.9\,pb.

\end{document}